\documentclass{article}

\usepackage{arxiv}

\usepackage{cite}
\usepackage[utf8]{inputenc}
\usepackage[T1]{fontenc}
\usepackage{hyperref}
\usepackage{url}
\usepackage{amsfonts}
\usepackage{nicefrac}
\usepackage{microtype}
\usepackage{lipsum}	
\usepackage{graphicx}
\usepackage{amsmath}

\title{Machine learning model to cluster and map tribocorrosion regimes in feature space}

\date{June 11, 2020}

\author{ \href{https://orcid.org/0000-0002-2668-1938}{\includegraphics[scale=0.06]{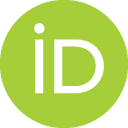}\hspace{1mm}Rahul Ramachandran}\\
	Department of Mechanical Engineering\\
	University of Nevada, Reno\\
	Reno, NV 89557 \\
	\texttt{r.ramacha6@gmail.com} \\
}

\hypersetup{
pdftitle={ML tribocorrosion},
pdfsubject={cs.LG, cond-mat.mtrl-sci},
pdfauthor={Rahul Ramachandran},
pdfkeywords={tribocorrosion maps, machine learning, SVM, K-means clustering},
}

\begin{document}
\maketitle

\begin{abstract}
Tribocorrosion maps serve the purpose of identifying operating conditions for acceptable rate of degradation. This paper proposes a machine learning based approach to generate tribocorrosion maps, which can be used to predict tribosystem performance. First, unsupervised machine learning is used to identify and label clusters from tribocorrosion experimental data. The identified clusters are then used to train a support vector classification model. The trained SVM is used to generate tribocorrosion maps. The generated maps are compared with the standard maps from literature.
\end{abstract}


\section{Introduction}
Development and implementation of predictive models from experimental data using machine learning (ML) have been an active research topic in material science and engineering. In ML, a computer program based on an algorithm learns from historical data to improve a performance metric. The selection of the algorithm, and metric depends on the specific data and problem at hand. The commonly used ML algorithms can be divided into regression, probability estimation, classification and clustering \cite{Liu2017}. Recently ML tools have been used in tribology to build predictive models for friction coefficient \cite{Gyurova2011,Xie2019}, wear rate \cite{Argatov2019} and wear volume \cite{Velten2000}, to design lubricants \cite{Bhaumik2019, Humelnicu2019,Rashmi2019} and functional materials \cite{Vinoth2020}, for tribocorrosion \cite{Pai2008} and surface roughness \cite{Buj-Corral2020} modeling, for wear particle classification \cite{Peng2019}, and in corrosion modeling \cite{Chelariu2015,Chou2017,Jiminez2015,Kamrunnahar2010,Mareci2015,Mareci2016,Xu1997}. \par
The combined effect of wear and corrosion, known as tribocorrosion, is often undesirable and can result in accelerated material degradation. It is frequently encountered where surfaces are in contact with each other in a corrosive environment. This phenomenon is relevant to a range of industries from energy and transportation to healthcare. Interfacial conditions such as materials, lubrication, normal load, relative speed, surface topography, temperature, humidity, pH etc. can affect the degradation process. The total mass change by tribocorrosion ($K_{wc}$) can be explained using the analysis in \cite{Yue1987} as

\begin{equation}
K_{wc} = K_{w} + K_{c}
\label{eq:eq1}
\end{equation}

where $K_{w}$ is the total mass of material removed due to wear, $K_{c}$ is the total mass of material removed due to corrosion. $K_{w}$ is $K_{wo} + \Delta K_{w}$, where $K_{wo}$ is the mass loss by wear in the absence of corrosion, and $\Delta K_w$ is the synergistic effect of corrosion on wear. $K_c$ is $K_{co} + \Delta K_c$, where $K_{co}$ is the mass loss by corrosion in the absence of wear, and $\Delta K_c$ is the additive effect (enhancement) of corrosion due to wear. These terms can be determined experimentally and used to characterize the dominant degenerative mechanism for the tribosystem. Tribocorrosion regimes \cite{Stack1996} are defined based on the ratio $K_c/K_w$ as follows:

\begin{equation}
K_{c}/K_{w} \leq 0.1, \qquad \text{wear}
\label{eq:eq2}
\end{equation}
\begin{equation}
0.1< K_{c}/K_{w} \leq 1, \qquad \text{wear-corrosion}
\label{eq:eq3}
\end{equation}
\begin{equation}
1 < K_{c}/K_{w} \leq 10, \qquad \text{corrosion-wear}
\label{eq:eq4}
\end{equation}
\begin{equation}
K_{c}/K_{w} > 10, \qquad \text{corrosion}
\label{eq:eq5}
\end{equation}

Tribocorrosion mechanism maps and synergy maps enable to illustrate these regimes as functions of interfacial conditions such as normal load, sliding speed, pH etc. These maps are helpful to identify conditions of minimal degradation, for process optimization, and in the design of functional materials and coatings \cite{Wood2007}. \par
The standard approach to create tribocorrosion maps involve extensive experiments to generate data, determination of the regimes, and creating graphical illustrations. Machine learning can be used to create accurate predictive models from the experimental data quickly and automatically. Although ML tools are widely used in tribology, no previous studies have dealt with identifying clusters and creating tribocorrosion maps using ML tools. This study uses both unsupervised and supervised learning techniques to make predictive models for tribocorrosion and to generate maps.

\section{Proposed Approach}
The predictive model is developed in two parts. The first part involves identifying clusters in tribocorrosion experimental data using the K Means clustering technique. The second part involves training a support vector classifier using the dataset and the cluster labels. The support vectors can then be used to draw wear-corrosion mechanism, synergy, and wastage maps. Two experimental datasets from published literature are used in this study. Dataset 1 is tribocorrosion data of Ti–25Nb–3Mo–3Zr–2Sn alloy in simulated Hank’s solution \cite{Wang2016}, and dataset 2 is tribocorrosion data of Co–Cr/UHMWPE couple in Ringer’s solution \cite{Stack2010}. These materials are relevant because of their applications in the biomedical industry for manufacturing implants. It is essential to understand the degradation and predict lifecycle of implants under various conditions, both of which can possibly be done with ML. The data was standardized by removing the mean and scaling to unit variance before training the model. The models were developed using the open-source ML package scikit-learn \cite{Pedregosa2011}. \par
\subsection{Unsupervised learning}
Clustering is a multivariate statistical technique used to group data into clusters based on their underlying structure. A commonly used technique is the K-means clustering. The number of clusters (K) and the starting centroids are provided as input parameters. This is followed by an iterative process of assigning the data points to each cluster based on their Euclidean squared distance to the corresponding centroid, and recalculating the centroids, until the centroids can no longer be adjusted. \par
In this study K-means clustering was used for exploratory analysis. For this, an unsupervised ML model was developed using the standardized datasets. In order to determine the optimal value of K for each dataset, the elbow method \cite{Milligan1985} and the silhouette coefficients \cite{Rousseeuw1987} were used.
\subsection{Supervised learning}
Support Vector Machines (SVMs) \cite{Cortes1995} are supervised ML algorithms that can be used in classification and regression \cite{Raghavendra2014}. SVMs use kernel functions to map datapoints from an input space to a high dimensional feature space to establish linear decision boundaries (hyper planes). These boundaries become nonlinear when transformed back to original input space, thus making non-linear classification possible. The datapoints closest to the hyperplane are called support vectors. Four basic kernel functions $K(x_i,x_j)$ are linear (eq.~\ref{eq:eq6}), polynomial (eq.~\ref{eq:eq7}), radial basis function (RBF) (eq.~\ref{eq:eq8}) and sigmoid (eq.~\ref{eq:eq9}). The hyperparameters $\gamma, r,$ and $d$ associated with the SVMs are generally tuned using cross-validation or grid search \cite{Hsu2003}.
\begin{equation}
K(x_i,x_j) = x_i^T x_j
\label{eq:eq6}
\end{equation}

\begin{equation}
K(x_i,x_j) = (\gamma x_i^T x_j + r)^d, \gamma>0
\label{eq:eq7}
\end{equation}

\begin{equation}
K(x_i,x_j) = e^{-\gamma \|x_i - x_j\|^2}, \gamma>0
\label{eq:eq8}
\end{equation}

\begin{equation}
K(x_i,x_j) = \tanh{(\gamma x_i^T x_j + r)}, \gamma>0
\label{eq:eq9}
\end{equation}
The labelled datasets obtained after clustering is used to train SVM classification models. Since SVMs are designed for binary classification, different strategies can be adopted based on the multiclass classification problem at hand \cite{Brownlee2020}.

\section{Results and Discussion}
\subsection{Identifying the clusters}
\begin{figure}[htbp]
\begin{center}
\includegraphics[width=6.5in]{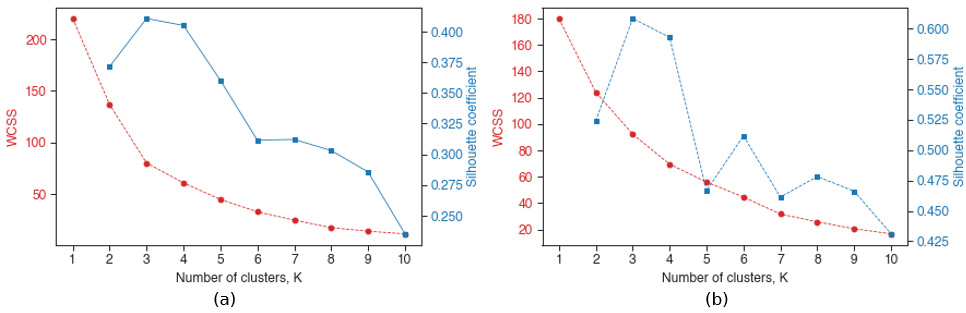}
\caption{Metrics to determine the optimum number of clusters for (a) dataset 1 and (b) dataset 2. Within-cluster sum of squares (WCSS) is denoted by the red circles and dashed line. Silhouette coefficients are denoted by blue squares and solid line.}
\label{fig:fig1}
\end{center}
\end{figure}

Clustering was done for a range of K values based on the data, and the resulting within-cluster sum of squares (WCSS) was obtained. The red lines in Figures \ref{fig:fig1}a and \ref{fig:fig1}b show the WCSS for datasets 1 and 2 respectively. The elbow can be observed at K=3 in Figure \ref{fig:fig1}a. However, the elbow in Figure \ref{fig:fig1}b is not obvious. Further analysis was done by calculating the silhouette coefficients. The blue lines in Figures \ref{fig:fig1}a and 1\ref{fig:fig1}b show the silhouette coefficients for a range of K values for datasets 1 and 2 respectively. The best value of K is where the silhouette coefficient is maximum, which is equal to 3 for both the datasets. \par

The two datasets were clustered and labelled using K-means clustering with K=3. The three clusters in dataset 1 can clearly be distinguished on a $K_w$ versus $K_c$ plot as shown in Figure \ref{fig:fig2}a. The single datapoint that forms a distinct cluster is identified as the corrosion-wear cluster, as it has a high $K_c$ compared to $K_w$. The other two clusters are closer to the $K_w$ axis, denoting wear-corrosion synergy. The cluster near the origin is labelled as the low wear-corrosion synergy cluster, and the cluster on the right is labelled as the high wear-corrosion synergy cluster. \par
Similarly, the three clusters in dataset 2 can be clearly identified on a $K_c$ versus $K_w$ plot as shown in Figure \ref{fig:fig3}a. The cluster with the very small values of $K_c$, and large values of $K_w$ can be identified as having wear as the predominant degradation mechanism. The cluster with similar values of $K_c$ and $K_w$ can be identified as having corrosion-wear synergy. The third cluster approximately in the center of the plot, is classed as having wear-corrosion synergy. This classification is similar to the results obtained in the source \cite{Stack2010}.

\subsection{Tribocorrosion maps}
\begin{figure}[htbp]
\begin{center}
\includegraphics[width=6.5in]{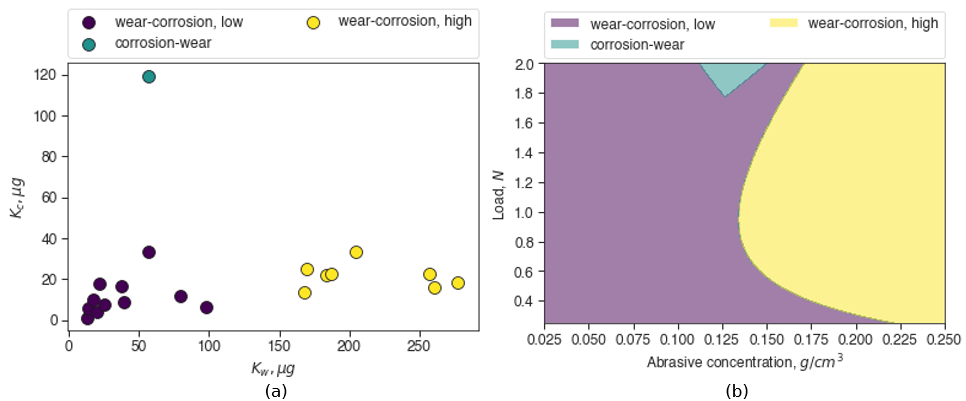}
\caption{(a) Three clusters in dataset 1 identified using K-means clustering (b) tribocorrosion mechanism map using SVM for the titanium alloy.}
\label{fig:fig2}
\end{center}
\end{figure}

Since the two datasets under consideration have three clusters each, a one-vs-one strategy is used to train the SVM and perform classification. The clusters in dataset 1 were labelled 0, 1 and 2. The features used were abrasive concentration ($g/cm^3$) and normal load ($N$). A polynomial kernel function was chosen to map the datapoints to a higher dimensional space. Values for the hyperparameters in eq.~\ref{eq:eq7} were as follows: kernel parameter $\gamma$ = 2, penalty parameter = 10, degree d = 2 and r = 0. The model was trained on the labelled dataset. The tribocorrosion map (Figure \ref{fig:fig2}b) of the feature space—abrasive particle concentration vs normal load—is generated by predicting using the trained model. The map shows the three previously identified clusters. From the map, the major degradation mechanism is identified as wear-induced corrosion, which is similar to the results obtained by \cite{Wang2016}. Corrosion-induced wear is the degradation mechanism for a small segment of the operating conditions. \par
\begin{figure}[htbp]
\begin{center}
\includegraphics[width=6.5in]{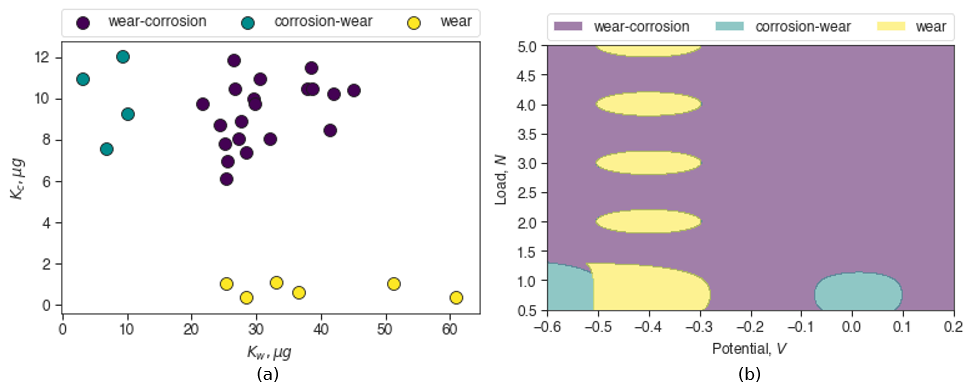}
\caption{(a) Three clusters in dataset 2 identified using K-means clustering (b) tribocorrosion mechanism map of Co-Cr generated using SVM.}
\label{fig:fig3}
\end{center}
\end{figure}

Similarly, the clusters in dataset 2 were labelled as 0, 1 and 2. The features used were potential (V) and normal load (N). RBF kernel (eq.~\ref{eq:eq8}) was used to map the dataset 2 to a higher dimensional space. The model was trained on the labelled dataset with hyperparameter $\gamma$ = 7 and the penalty parameter as 10. The tribocorrosion map obtained by predicting using the trained model on the feature space is shown in Figure \ref{fig:fig3}b. The three degradation mechanisms—wear, wear-corrosion and corrosion-wear—are mapped on the potential vs normal load graph. The wear-corrosion cluster as identified in Figure \ref{fig:fig3}a coves most of the area in the tribocorrosion map. Wear is dominant in the potential range -0.5 to -0.3 V. Corrosion-induced wear is dominant around potential of 0 V. \par

For the two tribosystems under consideration, the degradation mechanisms can be determined and thus an acceptable material loss rate can be maintained by controlling the interface conditions as per Figures \ref{fig:fig2}b and \ref{fig:fig3}b. Although the material data used in this study are from biomedical implant materials, the method proposed is relevant to applications such prediction of degradation mechanism, material selection, and lifecycle estimation of mining tools, automotive and marine machine components. \par
In this study the performance of the models was not evaluated using test and validation datasets due to limited sample sizes. Note that these are necessary to avoid over-fitting the models. In future work, more robust ML predictive models will be created using larger training dataset.

\section{Conclusion}
A machine learning based approach is proposed to generate tribocorrosion maps. First, tribocorrosion experimental data is clustered and labelled based on their underlying structure. The labelled datasets are used to train SVM classification models, with the labels being the targets. The problem being non-linear, RBF and polynomial kernels are used to map the data and establish the decision boundaries. The trained models are used to predict the targets on the feature space to generate the tribocorrosion maps. The tribocorrosion maps thus generated are relevant to applications such as material selection, identifying degradation mechanisms, and lifecycle estimation \par.


\end{document}